\documentclass{iopart}
\usepackage{psfig,epsfig,graphicx,float,pst-all,amssymb}
\renewcommand{\=}{~=~}
\newcommand{\be}{\begin{equation}}
\newcommand{\ee}{\end{equation}}
\newcommand{\bc}{\begin{center}}
\newcommand{\ec}{\end{center}}
\begin{document}

\title[$\beta-$soft, $\gamma-$soft rotor]{New analytic solutions of the 
collective Bohr hamiltonian for a $\beta-$soft, $\gamma-$soft axial rotor}
\author{Lorenzo Fortunato\footnote{\tt email: fortunat@pd.infn.it} 
and Andrea Vitturi}

\address{Dipartimento di Fisica ``G.Galilei'' and INFN, via Marzolo 8,
I-35131 Padova, Italy.}

\begin{abstract}
New analytic solutions of the quadrupole collective Bohr hamiltonian 
are proposed, exploiting an approximate separation of the $\beta$ and 
$\gamma$ variables to describe $\gamma-$soft prolate axial rotors. 
The model potential is a sum of two terms: a $\beta-$dependent term taken 
either with a Coulomb-like or a Kratzer-like form, and a $\gamma-$dependent 
term taken as an harmonic oscillator. 
In particular it is possible to give a one parameter
paradigm for a $\beta-$soft, $\gamma-$soft axial rotor that can be applied,
with a considerable agreement, to the spectrum of $^{234}$U.
\end{abstract}
\maketitle

\section{Introduction}
\begin{figure}[!t]
\begin{center}
\epsfig{file=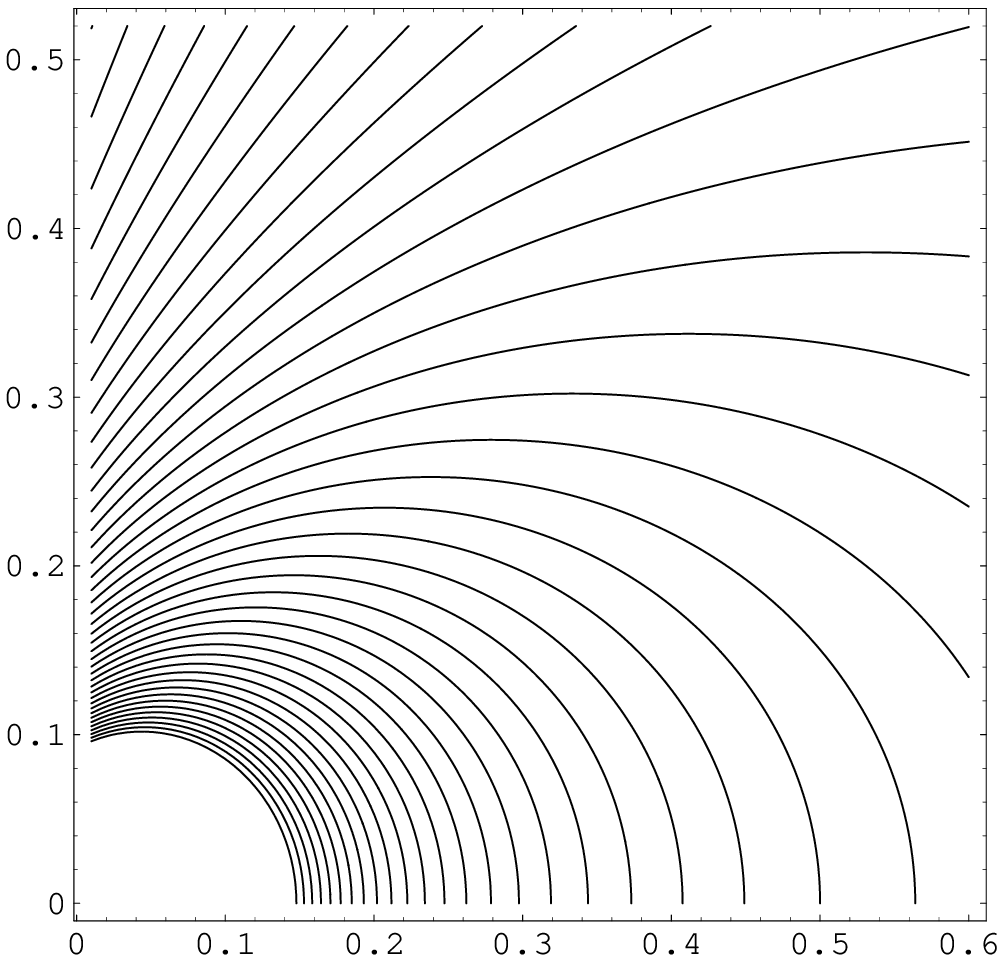,width=0.35\textwidth}~~~~
\epsfig{file=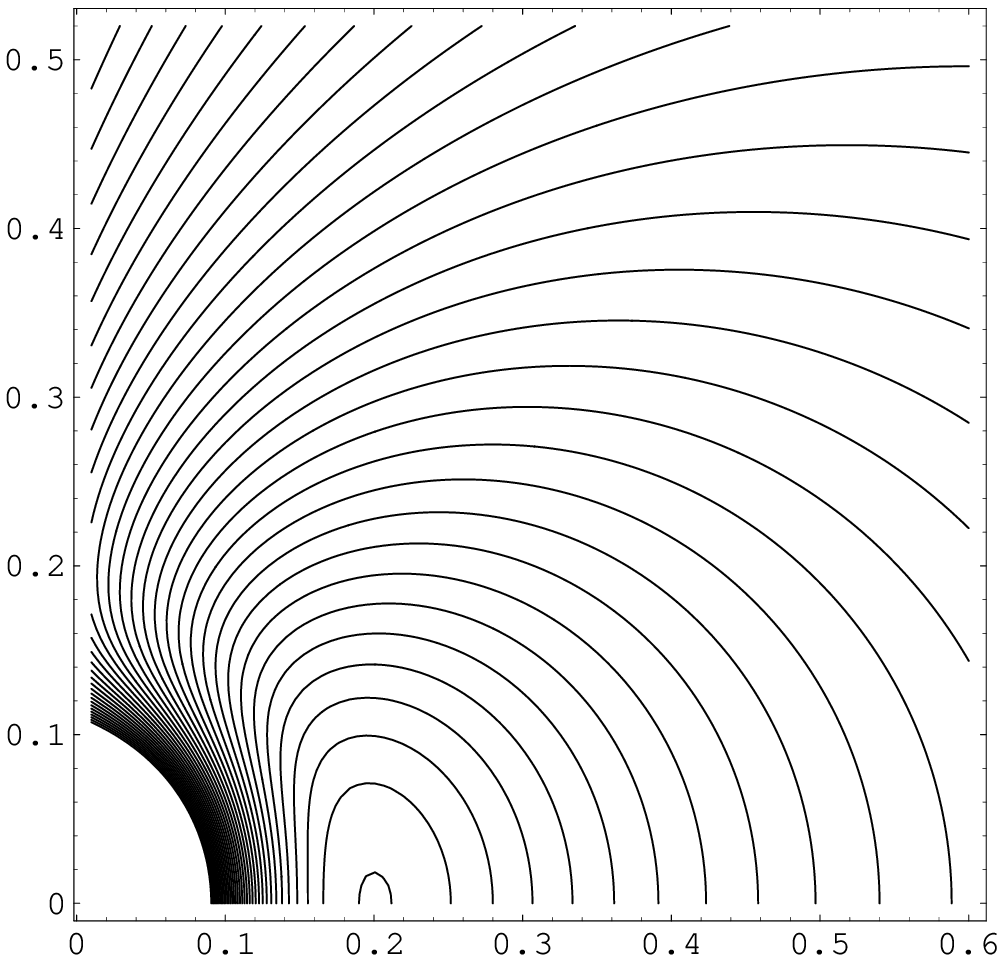,width=0.35\textwidth}
\end{center}
\vspace{-4.2cm}
\begin{center}
\begin{picture}(260,100)(0,0)
\psset{unit=1.pt}
\pspolygon*[linecolor=white,fillcolor=white,fillstyle=solid](3,6)
(69,122)(3,122)
\psline{-}(3,6)(69,122)
\psline{-}(3,6)(120,6)
\pspolygon*[linecolor=white,fillcolor=white,fillstyle=solid](150,6)
(216,122)(150,122)
\psline{-}(150,6)(216,122)
\psline{-}(150,6)(266,6)
\pswedge*[linecolor=black,fillcolor=black,fillstyle=solid](150,6){24}{0}{60}
\rput(-15,60){\tiny Y}
\rput(60,-10){\tiny X}
\rput(207,-10){\tiny X}
\rput{60}(33,70){\tiny $\gamma =60^o$ }
\rput{60}(180,70){\tiny $\gamma =60^o$ }
\rput(20,118){\tiny Coulomb}
\rput(166,118){\tiny Kratzer}
\end{picture}
\end{center}

\caption{Contour plots of the Coulomb-like and Kratzer-like potentials
 in $\beta$ plus harmonic
oscillator in $\gamma$. The coordinates are $X=\beta\cos{\gamma}$ and 
$Y=\beta\sin{\gamma}$. In the point $\beta_0=0$ in the Coulomb-like case 
the potential tends to $-\infty$, while in the the Kratzer-like case 
tends to $+\infty$ (black area).
The minimum in the right panel corresponds to $\beta_0=0.2$ and 
${\mathit D}=6$. }
\label{mappa_c}  
\end{figure}

We have recently discussed \cite{Fort} a new analytic solution of the 
differential equation $H\Psi\=E\Psi$, where $H$ is the Bohr hamiltonian with a 
$\gamma -$unstable class of potentials (called Coulomb-like and Kratzer-like
potentials), in the very same spirit of the solution given in ref. \cite{Iac1}
for the infinite square well. 
The hamiltonian discussed in the cited work displayed 
the dynamical symmetry SO(2,1)$\times$SO(5) 
$\supset$ SO(5)$\supset$ SO(3) $\supset$ SO(2).\\
Now we would like to take a step further and exploit the approximate
solution proposed in ref. \cite{Iac2} for the axially symmetric deformed rotor,
 replacing the infinite square well with our potentials for the 
$\beta$ part of the problem. This is a mere exercise in the Coulomb-like case,
because the peculiar behaviour of the $\beta$ potential around zero is rather
unrealistic in the nuclear case. The situation is more promising in the case 
of the Kratzer-like potential which displays a pocket with a deformed minimum
and therefore, combined with an harmonic dependence on the $\gamma$ variable,
may be taken as representative of $\beta-$soft, $\gamma-$soft rotor. 
For the sake of illustration we give in {fig. \ref{mappa_c}} two contour 
plots of the potential surfaces in
$\{\beta,\gamma\}$ plane resulting from the two cited $\beta$ 
potentials with an harmonic oscillator dependence on the $\gamma$ variable.
From symmetry considerations \cite{Bri} it is obvious that the $\gamma$ 
potential is restricted to be a function of $\cos{3\gamma}$. The choice
of the harmonic oscillator is thus meaningful only around $\gamma =0^o$, where
it can be regarded as the first order Taylor approximation of the correct
periodic even function.

Following the procedure of ref. \cite{Iac2}, expanding the moments of inertia
around $\gamma\=0^o$, the Bohr hamiltonian may be recast in the following form
$$      
\Biggl\{-{1\over \beta^4}{\partial \over \partial \beta}
\beta^4{\partial \over \partial \beta}-{1\over \beta^2\sin{3\gamma}}
{\partial \over \partial \gamma}\sin{3\gamma}{\partial \over \partial \gamma}
+{1\over 4\beta^2}\Biggl[{4\over 3}L(L+1)+K^2\Bigl({1\over \sin{\gamma}^2}-
{4\over 3}\Bigr) \Biggr] +$$\be
+u(\beta,\gamma)\Biggr\} \varphi_K^L(\beta,\gamma)= \varepsilon ~ 
\varphi_K^L(\beta,\gamma)
\label{b}
\ee
having expressed the eigensolutions as a product of a $(\beta,\gamma)$ part and
of an angular part (Wigner function) $\Psi(\beta,\gamma,\theta_i)\=
\varphi_K^L (\beta,\gamma){\cal D}_{M,K}^L(\theta_i)$ and having used the 
fact that within this approximation $K$ is a good quantum number.
As usual, reduced potential and energies, $u={2B_m\over \hbar^2}V$ and
$\varepsilon={2B_m\over \hbar^2}E$, have been introduced, being $B_m$ the
mass parameter of the Bohr hamiltonian.\\
Under the assumptions that the potential is written as
\be 
u(\beta,\gamma)\= u(\beta)+v(\gamma)
\label{poten}
\ee
and the $\gamma$ part has a 
minimum around $\gamma \= 0^o$, the equation admits an approximate 
separation of variables with the factorized solution 
$\varphi_K^L (\beta,\gamma)\=\xi_L(\beta)\eta_K(\gamma)$:
\be
\Biggl[  -{1\over \beta^4}{\partial \over \partial \beta}
\beta^4{\partial \over \partial \beta}+{1\over 4\beta^2}{4\over 3}L(L+1)
+u(\beta)\Biggr]\xi_L(\beta)=\varepsilon_\beta \xi_L(\beta)
\label{betapart} 
\ee
$$
\Biggl[ -{1\over \langle\beta^2\rangle \sin{3\gamma}} 
{\partial \over \partial \gamma}\sin{3\gamma}{\partial \over \partial \gamma}+
$$
\be
+{1\over 4\langle\beta^2\rangle} K^2\Bigl({1\over \sin{\gamma}^2}-{4\over 3}
\Bigr)+v(\gamma)\Biggr]\eta_K(\gamma) =\varepsilon_\gamma\eta_K(\gamma)
\label{gammapart}
\ee
with $\varepsilon \simeq \varepsilon^{(\beta)}+\varepsilon^{(\gamma)}$ and
where $\langle \beta^2 \rangle$ is the average of $\beta^2$ over $\xi(\beta)$.

In this note we will study the $\beta$ part of the problem (in Section 2) 
with different choices of
potentials, taking advantage (in Section 3) of Iachello's approximate 
solution of the equation in the $\gamma-$variable.
Electromagnetic transition rates are evaluated (in Section 4).

The peculiar feature of the present approach lies in the fact 
that the position 
of the ground state band and of all the $\beta-$bands, as well as
the corresponding moment of inertia along the bands, are fixed by 
only one parameter: an application of the solution with the 
Kratzer-like potential to the study of the spectrum of $^{234}$U is given 
(in Section 5).

\section{$\beta-$part of the problem}
With the substitution $\xi_L(\beta)\= \chi_L(\beta)\beta^{-2}$,
equation (\ref{betapart}) may be simplified to its standard form:
\be
{\partial^2 \over \partial \beta^2}\chi_L(\beta) +\Biggl\{ 
\varepsilon^{(\beta)}
-u(\beta)-\Biggl({2+{L(L+1)\over 3} \over \beta^2} \Biggr)\Biggr\}\chi_L(\beta)
\= 0
\label{general}
\ee
The solution of this second order differential equation yields the 
eigenfunctions in the $\beta$ variable and the eigenenergies $\varepsilon^{_
(\beta)}$. This is not the full solution of the problem, that requires also the
solution of the equation in the $\gamma-$variable (\ref{gammapart}), 
but represents the
subset of the full spectrum corresponding to zero quanta in the 
$\gamma$ degree of freedom and lying lower in energy.  
In this section we restrict our discussion to the $\beta-$part, drawing a 
parallel between two cases corresponding to
the two potentials, the Coulomb-like and 
the Kratzer-like potential, introduced in \cite{Fort}, leaving the 
considerations upon the $\gamma-$part of the problem for the next section.

\subsection{Coulomb-like case}

\begin{figure}[!t]
\bc
\vspace{0.6cm}
\begin{picture}(230,300)(0,0)
\psset{unit=1.5pt}
\psline[linewidth=0.5]{->}(0,20)(0,210)
\psline[linewidth=0.5]{-}(0,20)(1,20)\rput(-6,20){\tiny $0.0$}
\psline[linewidth=0.5]{-}(0,50)(1,50)\rput(-6,50){\tiny $0.5$}
\psline[linewidth=0.5]{-}(0,80)(1,80)\rput(-6,80){\tiny $1.0$}
\psline[linewidth=0.5]{-}(0,110)(1,110)\rput(-6,110){\tiny $1.5$}
\psline[linewidth=0.5]{-}(0,140)(1,140)\rput(-6,140){\tiny $2.0$}
\psline[linewidth=0.5]{-}(0,170)(1,170)\rput(-6,170){\tiny $2.5$}
\psline[linewidth=0.5]{-}(0,200)(1,200)\rput(-6,200){\tiny $3.0$}
\rput(55,5){\small $v=0$}
\psline{-}(40,20)(70,20)
\rput(25,20){\tiny $0$}\rput(45,23){\tiny $L=0$}
\psline{-}(40,80)(70,80)
\rput(25,80){\tiny $1$}\rput(45,83){\tiny $L=2$}
\psline{-}(40,123.1)(70,123.1)
\rput(25,123.1){\tiny $1.718$}\rput(45,126.1){\tiny $L=4$}
\psline{-}(40,143.75)(70,143.75)
\rput(25,143.75){\tiny $2.062$}\rput(45,146.75){\tiny $L=6$}
\psline{-}(40,154.25)(70,154.25)
\rput(25,154.25){\tiny $2.238$}\rput(45,157.25){\tiny $L=8$}
\psline[linestyle=dashed]{-}(40,175.7)(200,175.7)
\rput(25,175.7){\tiny $2.562$}\rput(45,178.7){\tiny $L=\infty$}
\psline{->}(45,78)(45,26)\rput(55,50){\tiny \it 1.00}
\psline{->}(45,121)(45,86)\rput(55,103){\tiny \it 2.95}
\psline{->}(45,141)(45,129)\rput(55,135){\tiny \it 9.71}

\rput(120,93){\small $v=1$}
\psline{-}(105,105.4)(135,105.4)
\rput(90,105.4){\tiny $1.423$}\rput(110,108.4){\tiny $L=0$}
\psline{-}(105,125.23)(135,125.23)
\rput(90,125.23){\tiny $1.754$}\rput(110,128.23){\tiny $L=2$}
\psline{-}(105,143.15)(135,143.15)
\rput(90,143.15){\tiny $2.052$}\rput(110,146.15){\tiny $L=4$}
\psline{-}(105,153.6)(135,153.6)
\rput(90,153.6){\tiny $2.227$}\rput(110,156.6){\tiny $L=6$}
\psline{->}(110,123)(110,111)\rput(120,117){\tiny \it 4.59}
\psline{->}(110,141)(110,131)\rput(120,136){\tiny \it 7.00}

\rput(185,125.4){\small $v=2$}
\psline{-}(170,135.27)(200,135.27)
\rput(155,135.27){\tiny $1.921$}\rput(175,138.27){\tiny $L=0$}
\psline{-}(170,144.15)(200,144.15)
\rput(155,144.15){\tiny $2.069$}\rput(175,147.15){\tiny $L=2$}
\psline{-}(170,153.27)(200,153.27)
\rput(155,153.27){\tiny $2.221$}\rput(175,156.27){\tiny $L=4$}
\psline{->}(185,142)(185,137)\rput(195,140){\tiny \it 13.39}

\psline{->}(168,133)(137,127) \rput(148,125){\tiny \it 32.7}
\psline{->}(103,103)(72,82) \rput(82,84){\tiny \it 6.96}

\psline{->}(103,123)(72,24)\rput(83,35){\tiny \it 0.02}
\psline{->}(103,123)(72,84)\rput(77,97){\tiny \it 0.34}
\psline{->}(103,123)(72,123)\rput(80,119){\tiny \it 13.7}

\pscircle*[linecolor=white](90,105.4){6}
\rput(90,105.4){\tiny $1.423$}

\end{picture}
\caption{$\beta-$spectrum of the Coulomb-like case ($n_\gamma=0$). 
The vertical scale is expressed in units of the energy of the 
$(v=0,L=2)$ state. Some B(E2) values are displayed in units of the
lowest transition of the first band (numbers in italics). 
Notice the presence and energy of 
a threshold and the absence of degeneracy.}
\label{sp1}  
\ec
\end{figure}
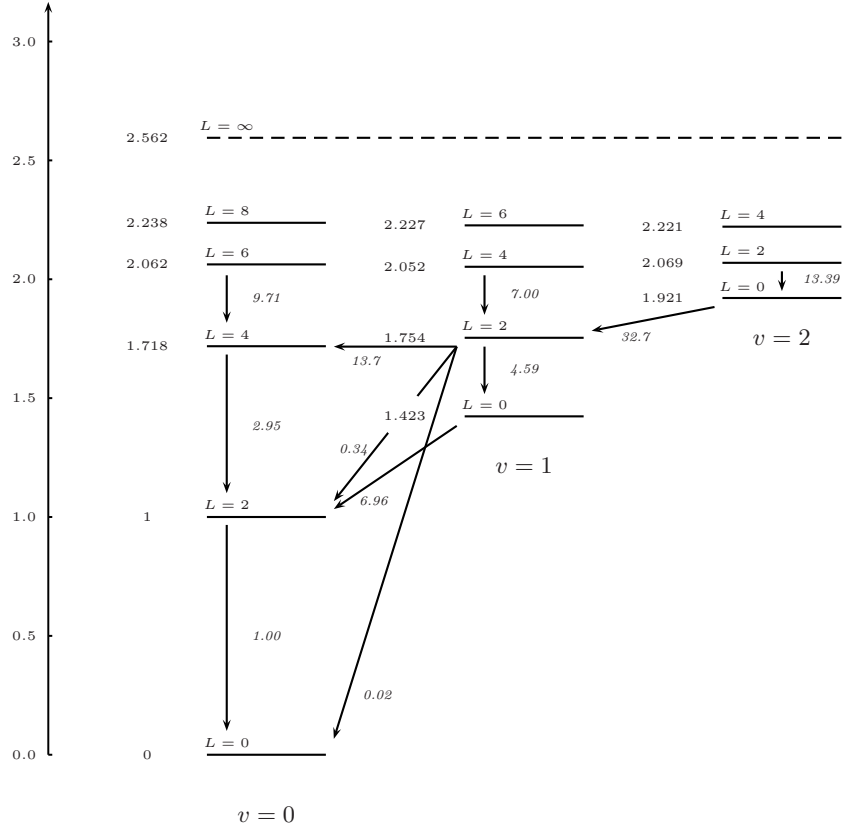

Inserting the Coulomb-like
potential $u(\beta)\=-A/\beta$ with $A>0$ and setting it in the variable
$x\=2\beta\sqrt{\epsilon}$, with the further substitutions 
$\varepsilon\=-\epsilon$,  $k={A\over2\sqrt{\epsilon}}$ and $\mu^2=
\bigl({9\over 4}+{L(L+1)\over 3}\bigr)$, we obtain the Whittaker's equation
\be
{\partial^2 \over \partial x^2}\chi_L(x) +\Biggl\{ -{1\over 4}
+{k\over x}+{1/4-\mu^2\over x^2} \Biggr\}\chi_L(x)\=0.
\ee
Its regular solution is the Whittaker function $M_{k,\mu}(x)$ that reads:
\be
\chi_L(x)\= {\cal N}e^{-x/2} x^{{1\over 2}+\mu} 
~_1F_1 \bigl({1\over 2}+\mu-k,1+2\mu,x\bigr)
\label{wave}
\ee 
which is, in general, a multivalued function. The constant ${\cal N}$ is 
determined from the normalization condition. We adopt usual conventions 
\cite{Wang} and thus the function is analytic on the real axis.
The hypergeometric series is an infinite series and to recover a good 
asymptotic behaviour we must require that it terminates, i.e. that it becomes 
a polynomial. This happens when the first argument is a negative integer, 
$-v$, that must thence be regarded as an additive quantum number. 
This condition fixes unambiguously the $\beta$ part of the spectrum:
\be
\epsilon_{v,L}^{(\beta)} \= 
{A^2/4 \over\Biggl(\sqrt{{9\over 4}+{L(L+1)\over 3}}+{1\over 2}+v \Biggr)^2}.
\label{spectr1}
\ee

Qualitatively this spectrum (fig. \ref{sp1}) looks rather similar to the 
one studied in the case of $\gamma-$instability. The states are grouped into
bands corresponding to different values of $v$; within each $v$ all the angular
momenta consistent with the constraints given by the $\gamma-$part (see 
Section 3) are present.
Note however that, fixing to unity the energy difference of the first two
states of the lowest band $(\epsilon_{0,2} -\epsilon_{0,0})$, 
the energy of first $4^+$ state is $1.718$, while the energy of the 
$(v=1,L=0)$ state (beginning of the second band) is $1.423$. 
These values should be compared with the energy ($1.35$) of the 
corresponding degenerate states of the $\gamma-$unstable case of 
ref. \cite{Fort}.
Other peculiarities are the presence 
of the threshold, corresponding to an infinite $L$ quantum number, that is 
located at $2.5616$ and the absence of the degeneracy pattern found in the 
$\gamma-$unstable case. Only accidental degeneracies may occur, but they 
are quite rare (the lowest is between the $L=14,v=0$ and $L=0,v=7$ states). \\

\begin{figure}[!t]
\bc
\begin{picture}(300,250)(25,0)
\psset{unit=0.9pt}
\psline[linewidth=0.5]{->}(0,20)(0,240)
\psline[linewidth=0.5]{-}(0,20)(1,20)\rput(-6,20){\tiny $0.0$}
\psline[linewidth=0.5]{-}(0,50)(1,50)\rput(-6,50){\tiny $0.5$}
\psline[linewidth=0.5]{-}(0,80)(1,80)\rput(-6,80){\tiny $1.0$}
\psline[linewidth=0.5]{-}(0,110)(1,110)\rput(-6,110){\tiny $1.5$}
\psline[linewidth=0.5]{-}(0,140)(1,140)\rput(-6,140){\tiny $2.0$}
\psline[linewidth=0.5]{-}(0,170)(1,170)\rput(-6,170){\tiny $2.5$}
\psline[linewidth=0.5]{-}(0,200)(1,200)\rput(-6,200){\tiny $3.0$}
\psline[linewidth=0.5]{-}(0,230)(1,230)\rput(-6,230){\tiny $3.5$}
\rput(55,5){\small $v=0$}
\psline{-}(40,20)(70,20)
\rput(25,20){\tiny $0$}\rput(45,23){\tiny $L=0$}
\psline{-}(40,80)(70,80)
\rput(25,80){\tiny $1$}\rput(45,83){\tiny $L=2$}
\psline{-}(40,123.1)(70,123.1)
\rput(25,123.1){\tiny $1.718$}\rput(45,126.1){\tiny $L=4$}
\psline{-}(40,143.75)(70,143.75)
\rput(25,143.75){\tiny $2.062$}\rput(45,146.75){\tiny $L=6$}
\psline{-}(40,154.25)(70,154.25)
\rput(25,154.25){\tiny $2.238$}\rput(45,157.25){\tiny $L=8$}
\psline[linestyle=dashed]{-}(40,173.72)(135,173.72)
\rput(25,173.72){\tiny $2.562$}\rput(45,176.72){\tiny $L=\infty$}
\rput(120,85){\small $v=1$}
\psline{-}(105,105.4)(135,105.4)
\rput(90,105.4){\tiny $1.423$}\rput(110,108.4){\tiny $L=0$}
\psline{-}(105,125.23)(135,125.23)
\rput(90,125.23){\tiny $1.754$}\rput(110,128.23){\tiny $L=2$}
\psline{-}(105,143.15)(135,143.15)
\rput(90,143.15){\tiny $2.052$}\rput(110,146.15){\tiny $L=4$}
\psline{-}(105,153.6)(135,153.6)
\rput(90,153.6){\tiny $2.227$}\rput(110,156.6){\tiny $L=6$}
\psline{->}(45,78)(45,25)\rput(55,50){\tiny \it 1.00}
\psline{->}(45,121)(45,85)\rput(55,103){\tiny \it 2.94}
\psline{->}(45,141)(45,129)\rput(55,135){\tiny \it 9.71}
\psline{->}(110,123)(110,110)\rput(120,117){\tiny \it 4.59}
\psline{->}(110,141)(110,131)\rput(120,136){\tiny \it 7.00}
\rput(90,230){\small $B=0.001$ }

\rput(190,5){\small $v=0$}
\psline{-}(175,20)(205,20)
\rput(160,20){\tiny $0$}\rput(180,23){\tiny $L=0$}
\psline{-}(175,80)(205,80)
\rput(160,80){\tiny $1$}\rput(180,83){\tiny $L=2$}
\psline{-}(175,133.04)(205,133.04)
\rput(160,133.04){\tiny $1.884$}\rput(180,136.04){\tiny $L=4$}
\psline{-}(175,161.84)(205,161.84)
\rput(160,161.84){\tiny $2.364$}\rput(180,164.84){\tiny $L=6$}
\psline{-}(175,177.4)(205,177.4)
\rput(160,177.4){\tiny $2.623$}\rput(180,180.4){\tiny $L=8$}
\psline[linestyle=dashed]{-}(175,207.8)(270,207.8)
\rput(160,207.8){\tiny $3.131$}\rput(180,210.8){\tiny $L=\infty$}
\rput(255,96.5){\small $v=1$}
\psline{-}(240,116.54)(270,116.54)
\rput(225,116.54){\tiny $1.609$}\rput(245,119.54){\tiny $L=0$}
\psline{-}(240,138.56)(270,138.56)
\rput(225,138.56){\tiny $1.976$}\rput(245,141.56){\tiny $L=2$}
\psline{-}(240,161.8)(270,161.8)
\rput(225,161.8){\tiny $2.363$}\rput(245,164.8){\tiny $L=4$}
\psline{-}(240,176.66)(270,176.66)
\rput(225,176.66){\tiny $2.611$}\rput(245,179.66){\tiny $L=6$}
\rput(225,230){\small $B=1$ }

\psline{->}(180,78)(180,25)\rput(190,50){\tiny \it 1.00}
\psline{->}(180,131)(180,85)\rput(190,107){\tiny \it 2.54}
\psline{->}(180,159)(180,138)\rput(190,150){\tiny \it 7.11}
\psline{->}(245,136)(245,121)\rput(255,128){\tiny \it 4.21}
\psline{->}(245,159)(245,143)\rput(255,151){\tiny \it 6.21}

\rput(325,5){\small $v=0$}
\psline{-}(310,20)(340,20)
\rput(295,20){\tiny $0$}\rput(315,23){\tiny $L=0$}
\psline{-}(310,80)(340,80)
\rput(295,80){\tiny $1$}\rput(315,83){\tiny $L=2$}
\psline{-}(310,219.98)(340,219.98)
\rput(295,219.98){\tiny $3.333$}\rput(315,222.98){\tiny $L=4$}
\rput(360,230){\small $B=10^6$ }
\psline{-}(370,200.)(400,200.)\psline{->}(385,200)(385,220)
\rput(375,203){\tiny $L=0$}\rput(390,220){\tiny $\infty$}
\rput(385,180){\small $v=1$}

\psline{->}(315,78)(315,25)\rput(325,50){\tiny \it 1.00}
\psline{->}(315,218)(315,85)\rput(325,151){\tiny \it 1.43}  
\end{picture}
\caption{
Evolution of the $\beta-$spectrum of the Kratzer-like rotor with
the parameter $B$. The vertical scale is expressed in units of the energy 
of the $(v=0,L=2)$ state. The B(E2) values, calculated with formula 
(\ref{be2}), between the lowest states are indicated in italics 
beside the corresponding downward arrows. Intraband transitions are 
not marked for the sake of simplicity.}
\label{sp2}  
\ec
\end{figure}
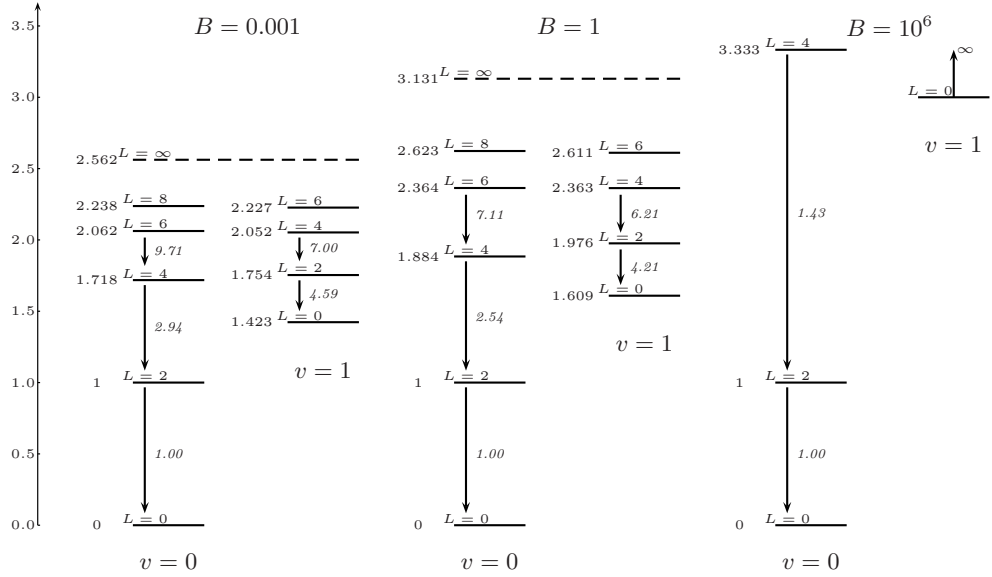

\subsection{Kratzer-like case}
We consider now the case of the Kratzer-like potential. 
Inserting in eq. (\ref{general}) the potential 
\be
u(\beta)=-A/\beta+B/\beta^2
\label{potenKr}
\ee 
and setting $x\=2\beta\sqrt{\epsilon}$, 
$\varepsilon\=-\epsilon$,  $k={A\over2\sqrt{\epsilon}}$ and $\mu^2=
\bigl(B+{9\over 4}+{L(L+1)\over 3}\bigr)$, 
we obtain again the Whittaker's equation whose
solutions may be written as in eq. (\ref{wave}) and we may repeat the 
whole procedure of the previous section, the only major difference being the 
definition of $\mu$. The spectrum assumes in this case the form
\be
\epsilon_{v,L}^{(\beta)} \= 
{A^2/4 \over\Biggl(\sqrt{B+{9\over 4}+{L(L+1)\over 3}}+{1\over 2}+v \Biggr)^2}
 ~.
\label{ep2}
\ee
We recall that the two parameters used here, $A$ and $B$, have an 
immediate translation into the position, $\beta_0$, and depth of 
the minimum of the potential, ${\cal D}$, being valid the following relations: 
$A\=2\beta_0{\cal D}$ and $B\=\beta_0^2{\cal D}$. Consequently $\beta_0\=2B/A$
and ${\cal D}\=A^2/(4B)$.

In fig. \ref{sp2} we show the excitation spectrum of the Kratzer-like 
potential in units of $(\epsilon_{0,2}-\epsilon_{0,0})$. This spectrum is 
clearly independent of $A$. We thus study its evolution 
with the parameter $B$. As expected, when $B$ is very small 
we recover the Coulomb-like spectrum
of fig. \ref{sp1}, while when $B$ becomes very large we recover a typical 
rotational spectrum, as can be showed either numerically or analytically 
(for large $B$, $(\epsilon_{0,L}-\epsilon_{0,0})\propto L(L+1)/(B^2)$).
Among the two limits the ratio ${\epsilon_{0,4}-\epsilon_{0,0}\over 
\epsilon_{0,2}-\epsilon_{0,0}}$ (relevant for a comparison with experimental
data) varies between the two extremes, $1.718$ and $10/3$.

\section{$\gamma-$part of the problem}
The approximate solution of the $\gamma-$part of the problem follows 
closely the derivation in \cite{Iac2}. In this reference
the differential equation (\ref{gammapart}) in the $\gamma$ variable  
with an harmonic oscillator dependence upon $\gamma$ may 
be approximatively solved by expanding all the sine 
functions in powers of $\gamma$. Thus the differential equation to be solved
is 
\be
\Biggl[ -{1\over \langle \beta^2\rangle} {1\over \gamma} {\partial \over 
\partial \gamma} \gamma {\partial \over \partial \gamma} +
{K^2\over 4\langle \beta^2\rangle} \Bigl( {1\over \gamma^2} -{4\over 3}\Bigr)
+v(\gamma) -\varepsilon^{(\gamma)} \Biggr] \eta_{K}(\gamma)\=0. 
\ee
Choosing 
\be
v(\gamma)\= c\gamma^2/2
\label{potengam}
\ee
we have
\be
\varepsilon^{(\gamma)}\={c \over \sqrt{\langle \beta^2 \rangle}} (n_\gamma+1) -
{K^2\over 3\langle \beta^2 \rangle}
\label{engam}
\ee
for eigenvalues, and 
\be
\eta_{n_\gamma,K}(\gamma)\= \gamma^{\mid K/2\mid} e^{-c\gamma^2/2}
L_n^{\mid K\mid} (c\gamma^2)
\ee
for eigenfunctions in terms of Laguerre polynomials, with 
$n=(n_\gamma-\mid K\mid)/2$ and $n_\gamma\=0,1,2,3,...~~$. A simple rule 
to express in a compact way the set of $K$ quantum numbers encountered for a 
given  $n_\gamma$ is
\be
K\= {2n_\gamma-4j}, \mbox{~~~ with  ~~~} j=0,\cdots ,n_\gamma ~.
\ee
The values of $L$ are determined by $K$: if $K=0$ then $L=0,2,4,...~~$, while
if $K\ne 0$ then $L=K,K+1,K+2,...~~$. 
Levels in fig. \ref{sp1} and
\ref{sp2}
correspond to $n_\gamma=0$, that implies $K=0$, and therefore only 
even values of $L$ are present. States labeled by the same set of 
quantum numbers, disregarding the sign of $K$, are degenerate in energy. 

\begin{figure}[!t]
\bc
\psset{xunit=12cm,yunit=2cm}
\begin{pspicture}(-0.05,6)(0.6,11)
\psline[linewidth=1pt]{->}(-0.05,10)(0.6,10)
\psline[linewidth=1pt]{->}(0,6)(0,11)
\psplot[plotstyle=curve,linewidth=1.2pt]{0.024}{0.08}{.01 x 2 exp div .4 x 
div sub 10 add}
\psplot[plotstyle=curve,linewidth=1.2pt]{0.08}{0.6}{.01 x 2 exp div .4 x 
div sub 10 add}
\rput(-0.02,10.8){\small $\varepsilon$}
\rput(0.55,10.1){\small $\beta$}
\rput(0.35,6.5){\small A=1600}
\rput(0.35,6.3){\small B=40}
\rput(0.29,6.4){\LARGE \{}
\rput(0.43,6.4){\LARGE \{}
\rput(0.5,6.5){\small $\beta_0$=0.05}
\rput(0.5,6.3){\small ${\cal D} \simeq$ -29 }

\rput(-0.05,6){\tiny -29}
\rput(-0.05,8.62){\tiny -10}
\rput(-0.05,7.24){\tiny -20}
\psline{-}(0.038,6.72)(0.085,6.72)\rput(0.06,6.77){\tiny 0,0} 
\psline{-}(0.037,6.86)(0.094,6.86)\rput(0.06,6.91){\tiny 0,2} 
\psline{-}(0.034,7.49)(0.13,7.49)\rput(0.06,7.535){\tiny 1,0} 
\psline{-}(0.034,7.59)(0.135,7.59)\rput(0.06,7.64){\tiny 1,2} 
\rput(0.25,10.5){\small $\gamma = 0^o$}
\end{pspicture}
\caption{Kratzer potential in $\beta$ for $\gamma =0$ in energy units of 
$(\epsilon_{0,2} -\epsilon_{0,0})$. The parameters 
(specified in the two equivalent sets on the figure) have been chosen
in such a way that the lowest states of the first two bands 
(labeled by their quantum numbers $v,L$) are deeply bound in the well.}
\label{caso}
\ec
\end{figure}
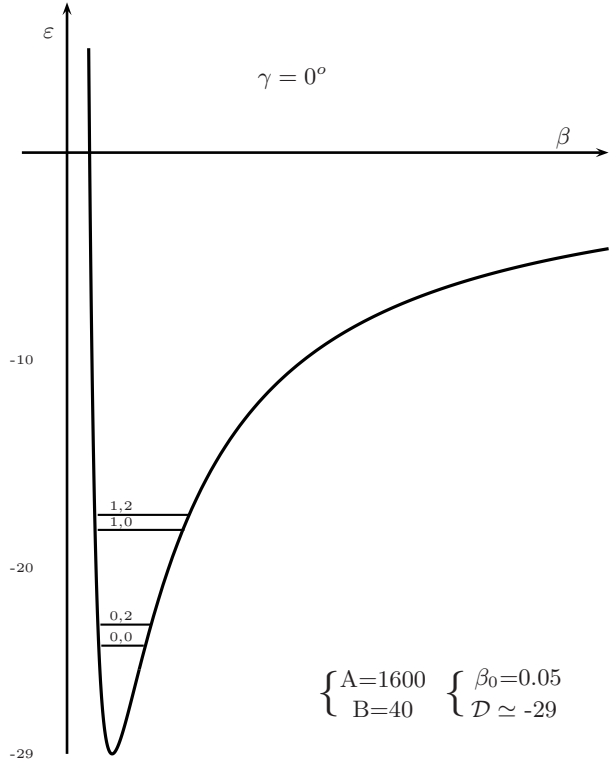

\section{Electromagnetic transitions}
The quadrupole electromagnetic operator reads 
\be
T_\mu = t \beta \Bigl[  {\cal D}^{(2)}_{\mu,0}(\Omega)\cos{\gamma}
+{1\over \sqrt{2}}\Bigl(  {\cal D}^{(2)}_{\mu,2}(\Omega)+
{\cal D}^{(2)}_{\mu,-2}(\Omega)\Bigr)\sin{\gamma} \Bigr] 
\ee
where $t$ is a scale factor. Limiting ourselves to 
transitions within the $n_\gamma=0$ bands and
exploiting the expansion of the trigonometric functions around $\gamma=0$,
transition rates may be evaluated
separating the integral in three parts: the angular integral is done 
analytically, exploiting the properties of the Wigner ${\cal D}$ functions, 
the integral in $\gamma$ reduces to an orthonormality condition, 
while the integral in $\beta$ has been calculated numerically employing the 
explicit forms of the wavefunctions,
$$
B(E2,v_iL_i \rightarrow v_f L_f )_{n_\gamma = 0}  =   $$
\be
{5\over 16 \pi} \langle L_i,0,2,0 \mid  L_f,0\rangle^2
\left| \int_0^\infty d\beta  \beta^5 \xi^*_{v_f,L_f}(\beta) 
\xi_{v_i,L_i}(\beta) \right|^2~. 
\label{be2}
\ee
The calculated $B(E2)$ values (normalized to the value for the transition 
from the first $2^+$ state to the $0^+$ ground state of the lowest band) 
are reported in fig. \ref{sp1} and \ref{sp2}. 
In the case of Kratzer potential, the B(E2) values tend to 
the Coulomb-like case values when the parameter $B$ is small and tend to 
the rigid rotor values when $B$ is large (for example in the case of the 
$4^+_0 \rightarrow 2^+_0$ transition we get the limiting value $10/7$).

\section{Discussion}
The $(\beta,\gamma)$ potential with the Kratzer form in the 
$\beta$ variable and the harmonic behaviour in $\gamma$
may be assumed as a representative of $\beta-$soft, $\gamma-$soft rotor.
\begin{figure}[!t]
\begin{center}
\vspace{11mm}
\epsfig{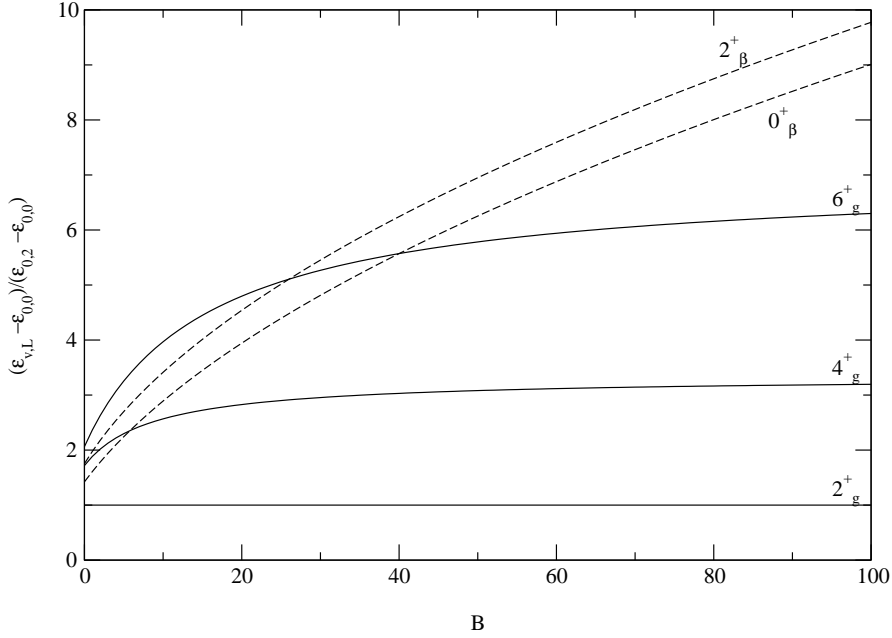}
\end{center}
\caption{Reduced energies normalized to the energy difference between
the first two states of the ground state band as a function of the parameter
$B$. The first three states of the g.s. band $(v=0)$ as well as the 
first two of the $\beta-$band $(v=1)$ are displayed. One can notice that 
the beginning of the 
second band lies at higher energies when the deformation of the system 
is increased and eventually tends to infinity in the rotor limit.}
\label{cro}
\end{figure}
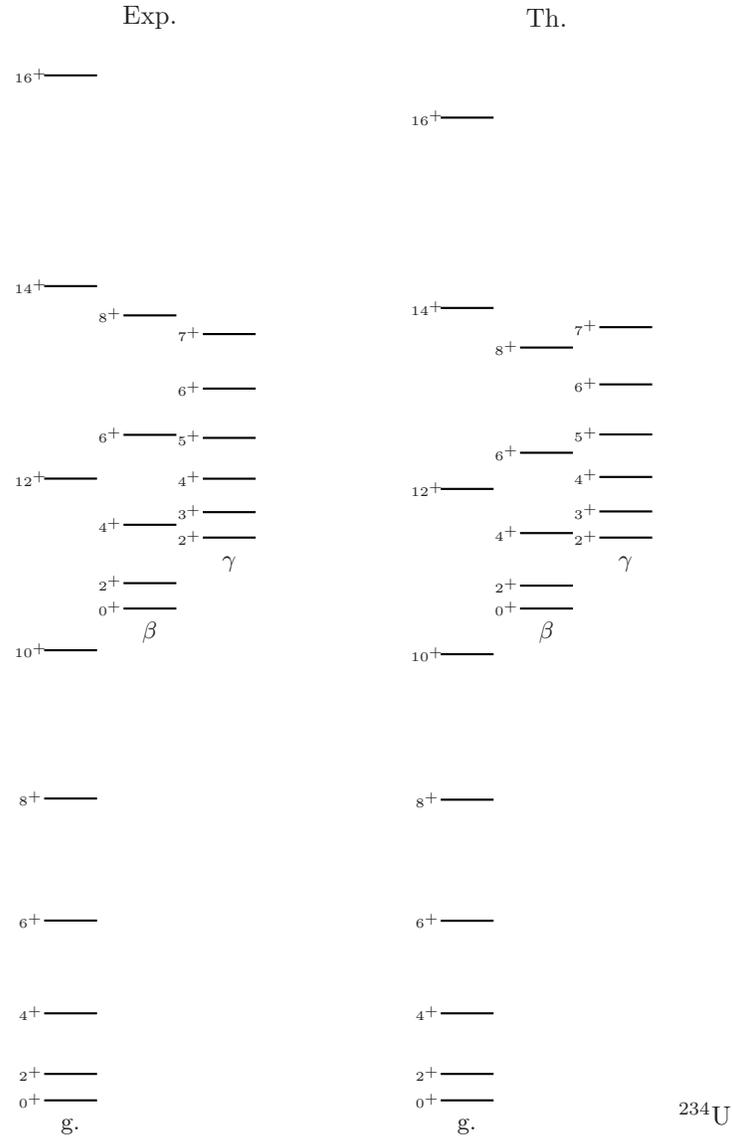
\begin{figure}[!t]
\bc
\begin{picture}(230,480)(0,0)
\psset{unit=1.pt}
\rput(60,430){Exp.}
\rput(30,10){\small g.}
\psline{-}(20,20)(40,20)\rput(15,20){\tiny $0^+$}
\psline{-}(20,30)(40,30)\rput(15,30){\tiny $2^+$}
\psline{-}(20,52.94)(40,52.94)\rput(15,52.94){\tiny $4^+$}
\psline{-}(20,88.06)(40,88.06)\rput(15,88.06){\tiny $6^+$}
\psline{-}(20,134.26)(40,134.26)\rput(15,134.26){\tiny $8^+$}
\psline{-}(20,190.39)(40,190.39)\rput(15,190.39){\tiny $10^+$}
\rput(60,197){\small $\beta$}
\psline{-}(50,206.18)(70,206.18)\rput(45,206.18){\tiny $0^+$}
\psline{-}(50,215.79)(70,215.79)\rput(45,215.79){\tiny $2^+$}
\psline{-}(50,237.89)(70,237.89)\rput(45,237.89){\tiny $4^+$}
\psline{-}(50,271.93)(70,271.93)\rput(45,271.93){\tiny $6^+$}
\psline{-}(50,317.15)(70,317.15)\rput(45,317.15){\tiny $8^+$}
\psline{-}(20,255.36)(40,255.36)\rput(15,255.36){\tiny $12^+$}
\psline{-}(20,328.23)(40,328.23)\rput(15,328.23){\tiny $14^+$}
\psline{-}(20,408.)(40,408.)\rput(15,408.){\tiny $16^+$}
\rput(90,223){\small $\gamma$}
\psline{-}(80,233.04)(100,233.04)\rput(75,233.04){\tiny $2^+$}
\psline{-}(80,242.67)(100,242.67)\rput(75,242.67){\tiny $3^+$}
\psline{-}(80,255.33)(100,255.33)\rput(75,255.33){\tiny $4^+$}
\psline{-}(80,270.78)(100,270.78)\rput(75,270.78){\tiny $5^+$}
\psline{-}(80,289.44)(100,289.44)\rput(75,289.44){\tiny $6^+$}
\psline{-}(80,310.08)(100,310.08)\rput(75,310.08){\tiny $7^+$}
\rput(210,430){Th.}
\rput(180,10){\small g.}
\psline{-}(170,20)(190,20)\rput(165,20){\tiny $0^+$}
\psline{-}(170,30)(190,30)\rput(165,30){\tiny $2^+$}
\psline{-}(170,52.94)(190,52.94)\rput(165,52.94){\tiny $4^+$}
\psline{-}(170,87.98)(190,87.98)\rput(165,87.98){\tiny $6^+$}
\psline{-}(170,133.82)(190,133.82)\rput(165,133.82){\tiny $8^+$}
\psline{-}(170,188.89)(190,188.89)\rput(165,188.89){\tiny $10^+$}
\rput(210,197){\small $\beta$}
\psline{-}(200,206.18)(220,206.18)\rput(195,206.18){\tiny $0^+$}
\psline{-}(200,214.85)(220,214.85)\rput(195,214.85){\tiny $2^+$}
\psline{-}(200,234.74)(220,234.74)\rput(195,234.74){\tiny $4^+$}
\psline{-}(200,265.15)(220,265.15)\rput(195,265.15){\tiny $6^+$}
\psline{-}(200,304.98)(220,304.98)\rput(195,304.98){\tiny $8^+$}
\psline{-}(170,251.46)(190,251.46)\rput(165,251.46){\tiny $12^+$}
\psline{-}(170,319.97)(190,319.97)\rput(165,319.97){\tiny $14^+$}
\psline{-}(170,392.02)(190,392.02)\rput(165,392.02){\tiny $16^+$}
\rput(240,223){\small $\gamma$}
\psline{-}(230,233.04)(250,233.04)\rput(225,233.04){\tiny $2^+$}
\psline{-}(230,242.94)(250,242.94)\rput(225,242.94){\tiny $3^+$}
\psline{-}(230,255.99)(250,255.99)\rput(225,255.99){\tiny $4^+$}
\psline{-}(230,272.08)(250,272.08)\rput(225,272.08){\tiny $5^+$}
\psline{-}(230,291.03)(250,291.03)\rput(225,291.03){\tiny $6^+$}
\psline{-}(230,312.69)(250,312.69)\rput(225,312.69){\tiny $7^+$}
\rput(270,15){$^{234}$U}
\end{picture}
\ec
\caption{Comparison between the calculated and experimental spectrum of 
$^{234}$U (positive parity states). The theoretical 
predictions are obtained according to (\ref{ep2}) and (\ref{engam}) 
with the choice $B=393.7415$ and $c=-41.5629$ to define the $(\beta,\gamma)$ 
potential in (\ref{potenKr}) and (\ref{potengam}). The ground state, 
$\beta$ and $\gamma$ bands are displayed. }
\label{U}
\end{figure}
For illustration we display in fig. \ref{caso} a section of a 
Kratzer potential  
along the line with $\gamma =0^o$. The parameters have been chosen
to show how the well can be shaped in order to have a rotational
behaviour in presence of a $\beta-$softness.
In fact the position of
some of the lowest states of the first two bands is quite deep in the well:
this means that the variation of the mean square deformation
will not change much among these states (so that the approximate separation 
of variables in eq. (\ref{b}) is {\it a posteriori} justified). 
Note that the present solution requires
only one parameter to fix the relative position of all the states 
in the lowest band and the position of
the so-called $\beta-$bands (the parameter $A$ in the potential does not
play any role in determining the scaled eigenvalues).
The position of the $\gamma-$band is fixed by the additional parameter $c$ 
in eq. (\ref{engam}). As usual the physical values of the spectrum are then
obtained by an overall scaling factor.\\
As we move to large deformations and deeper minima by increasing the 
parameter $B$, the position
of the $\beta-$bands with respect to the ground state band moves higher
in energy and a crossing occurs between the relative positions of states
in different bands, as one can see in fig. \ref{cro}.

We applied our form of the potential to describe the lowest
rotational bands in $^{234}$U. A comparison between the lowest positive parity 
experimental energy levels of $^{234}$U \cite{expU} and the predictions of the
model presented here are given in fig. \ref{U}. The agreement is good 
in spite of the fact that all the relative positions of the states in the 
ground state band and $\beta-$band have been 
fixed with only one parameter, $B$, that has been obtained fitting the 
energy of the $4_g^+$ level.
 
Note that, at variance with the experiment that seems to display almost
equal moments of inertia for the ground and $\beta-$band, the theory gives 
a slightly larger moment for the $\beta-$band. This can be understood 
from the expansion of the general formula (\ref{ep2}) in powers of 
$L(L+1)$. In leading order one obtains, for the energy differences within
a generic band (characterized by the quantum number $v$), the following
expression
\be
\epsilon_{v,L}-\epsilon_{v,0}\= L(L+1){A^2/24\over 
\sqrt{B+9/4} (\sqrt{B+9/4}+1/2+v)^3},
\ee
which leads to the mild dependence on $v$, discussed above.

In summary we have discussed a new solution of the Bohr hamiltonian with
the aim of describing a $\beta-$soft, $\gamma-$soft axial rotor. We have
given spectra and transition rates that may in principle encompass a broad 
range of situations and can be used to survey experimental data. 
We have discussed, as an example, the spectrum of $^{234}$U,
that may be considered as a ``quasi-rigid'' rotor  and that is accurately
described within the present formalism, making use of the solution with
a Kratzer-like potential.

\end{document}